\documentclass[pdftex,twocolumn,epjc3,a4paper]{svjour3}

\RequirePackage[T1]{fontenc}
\smartqed  
\usepackage{amsmath}
\usepackage{amssymb}
\RequirePackage{mathptmx}      
\RequirePackage{flushend}
\RequirePackage{mathrsfs}
\RequirePackage{dsfont}
\usepackage{units}
\usepackage{graphicx}
\usepackage{cite}
\usepackage{xcolor} 
\RequirePackage[colorlinks,citecolor=blue,urlcolor=blue,linkcolor=blue]{hyperref}
\usepackage[switch]{lineno}
\journalname{Eur. Phys. J. C}
\begin{document}

\author
{%
A.~H.~Abdelhameed\thanksref{addr1}\and
G.~Angloher\thanksref{addr1}\and
P.~Bauer\thanksref{addr1}\and
A.~Bento\thanksref{addr1,addr2}\and
E.~Bertoldo\thanksref{t1,e1,addr1} \and
C.~Bucci\thanksref{addr3}\and
L.~Canonica\thanksref{addr1}\and
A.~D'Addabbo\thanksref{addr3,addr10}\and
X.~Defay\thanksref{addr4}\and
S.~Di~Lorenzo\thanksref{addr3,addr10}\and
A.~Erb\thanksref{addr4,addr5}\and
F.~v.~Feilitzsch\thanksref{addr4}\and
N.~Ferreiro~Iachellini\thanksref{addr1}\and
S.~Fichtinger\thanksref{addr7}\and
A.~Fuss\thanksref{addr7,addr8}\and
P.~Gorla\thanksref{addr3}\and
D.~Hauff\thanksref{addr1}\and
J.~Jochum\thanksref{addr6}\and
A.~Kinast\thanksref{addr4}\and
H.~Kluck\thanksref{addr7,addr8}\and
H.~Kraus\thanksref{addr9}\and
A.~Langenk\"amper\thanksref{addr4}\and
M.~Mancuso\thanksref{t1,e2,addr1}\and
V.~Mokina\thanksref{addr7}\and
E.~Mondragon\thanksref{addr4}\and
A.~M\"unster\thanksref{addr4}\and
M.~Olmi\thanksref{addr3,addr10}\and
T.~Ortmann\thanksref{addr4}
C.~Pagliarone\thanksref{addr3,addr11}\and
L.~Pattavina\thanksref{addr4, addr10}\and
F.~Petricca\thanksref{addr1}\and
W.~Potzel\thanksref{addr4}\and
F.~Pr\"obst\thanksref{addr1}\and
F.~Reindl\thanksref{t1,e3,addr7,addr8}\and
J.~Rothe\thanksref{addr1}\and
K.~Sch\"affner\thanksref{addr3,addr10}\and
J.~Schieck\thanksref{addr7,addr8}\and
V.~Schipperges\thanksref{addr6}\and
D.~Schmiedmayer\thanksref{addr7,addr8}\and
S.~Sch\"onert\thanksref{addr4}\and
C.~Schwertner\thanksref{addr7,addr8}
M.~Stahlberg\thanksref{addr7,addr8}\and
L.~Stodolsky\thanksref{addr1}\and
C.~Strandhagen\thanksref{addr6}\and
R.~Strauss\thanksref{addr4}\and
C.~T\"urko\u{g}lu\thanksref{addr7,addr8}\and
I.~Usherov\thanksref{addr6}\and
M.~Willers\thanksref{addr4}\and
V.~Zema\thanksref{t1,e4,addr3,addr10,chalmers}\\
(The CRESST Collaboration)\\
and\\
M.~Chapellier\thanksref{addr12}\and
A.~Giuliani\thanksref{addr12,addr13}\and
C.~Nones\thanksref{addr14}\and
D.V.~Poda\thanksref{addr12,addr17}\and
V.N.~Shlegel\thanksref{addr18}\and
M.~Vel\'azquez\thanksref{addr15}\and
A.S.~Zolotarova\thanksref{addr14,addr16}
}

\institute
{%
Max-Planck-Institut f\"ur Physik, D-80805 M\"unchen, Germany 
\label{addr1} \and
INFN, Laboratori Nazionali del Gran Sasso, I-67100 Assergi, Italy 
\label{addr3} \and
Physik-Department and Excellence Cluster Universe, Technische Universit\"at M\"unchen, D-85748 Garching, Germany 
\label{addr4} \and
Eberhard-Karls-Universit\"at T\"ubingen, D-72076 T\"ubingen, Germany 
\label{addr6} \and
Institut f\"ur Hochenergiephysik der \"Osterreichischen Akademie der Wissenschaften, A-1050 Wien, Austria 
\label{addr7} \and
Atominstitut, Vienna University of Technology, A-1020 Wien, Austria 
\label{addr8} \and
Department of Physics, University of Oxford, Oxford OX1 3RH, United Kingdom 
\label{addr9} \and
also at: GSSI-Gran Sasso Science Institute, 67100, L'Aquila, Italy 
\label{addr10} \and
also at: Departamento de Fisica, Universidade de Coimbra, P3004 516 Coimbra, Portugal 
\label{addr2} \and
also at: Walther-Mei{\ss}ner-Institut f\"ur Tieftemperaturforschung, D-85748 Garching, Germany 
\label{addr5} \and
also at: Chalmers University of Technology, Department of Physics, SE-412 96 G\"oteborg, Sweden 
\label{chalmers} \and
also at: Dipartimento di Ingegneria Civile e Meccanica, Universit\`{a} degli Studi di Cassino e del Lazio Meridionale, I-03043 Cassino, Italy 
\label{addr11} \and
CSNSM, Univ. Paris-Sud, CNRS/IN2P3, Universit\'e Paris-Saclay, 91405 Orsay, France
\label{addr12} \and
DISAT, Universit\`a dell'Insubria, 22100 Como, Italy
\label{addr13} \and
IRFU, CEA, Universit\'{e} Paris-Saclay, F-91191 Gif-sur-Yvette, France
\label{addr14} \and
Univ. Grenoble Alpes, CNRS, Grenoble INP, SIMAP UMR 5266, 38000 Grenoble, France
\label{addr15} \and
Institute for Nuclear Research, 03028 Kyiv, Ukraine
\label{addr17} \and
Nikolaev Institute of Inorganic Chemistry, 630090 Novosibirsk, Russia.
\label{addr18} \and
now at: CSNSM, Univ. Paris-Sud, CNRS/IN2P3, Universit\'e Paris-Saclay, 91405 Orsay, France 
\label{addr16}
}
\thankstext[$\star$]{t1}{Corresponding author}
\thankstext{e1}{bertoldo@mpp.mpg.de}
\thankstext{e2}{mancuso@mpp.mpg.de}
\thankstext{e3}{florian.reindl@oeaw.ac.at}
\thankstext{e4}{vanessa.zema@gssi.it}

\title{First results on sub-GeV spin-dependent dark matter interactions with $^{7}$Li}

\maketitle
\begin{abstract}
In this work, we want to highlight the potential of lithium as a target for spin-dependent dark matter search in cryogenic experiments, with a special focus on the low-mass region of the parameter space.\\
We operated a prototype detector module based on a Li$_2$MoO$_4$ target crystal in an above-ground laboratory. Despite the high background environment, the detector sets competitive limits on spin-dependent interactions of dark matter particles with protons and neutrons for masses between \unit[0.8]{GeV/c$^2$} and \unit[1.5]{GeV/c$^2$}. 
\end{abstract}

\section{Introduction} \label{sec:introduction}

In recent decades a significant experimental effort has been dedicated to the direct search of dark matter by multiple experiments. Most searches have focused on the dark matter particle mass range between $\sim$\unit[10] {GeV/c$^2$} and $\sim$\unit[100]{GeV/c$^2$}\\~\cite{Undagoitia}, but recently an increasing interest points towards models involving lighter particles~\cite{Kaplan09, Boehm03, Feng08}. More emphasis is also being given to interactions between dark matter particles and ordinary matter beyond the classic spin-independent interactions~\cite{Fitzpatrick12,Aprile2018,Angloher2018,Akerib2018}. 
In this work we present a first investigation of spin-de\-pendent interactions in the low dark matter particle mass range using well established cryogenic detection technologies with a target crystal containing lithium. To our knowledge, lithium has previously been used only in~\cite{Miuchi02} and~\cite{Belli2012} for direct dark matter detection. \\
The cryogenic detector technology used for direct detection of dark matter has demonstrated to be ideal to probe spin-independent interactions in the low mass (\unit[$\lesssim10$]{GeV/c$^2$}) parameter space. Different target materials are used by various experiments: CRESST opted for CaWO$_4$~\cite{Abdelhameed2019}, EDELWEISS for germanium~\cite{Arnaud17}, and CDMS based its technology on both germanium and silicon~\cite{Agnese16}. However, this technology is not yet fully exploited to investigate spin-depen\-dent interactions in the very same region of the parameter space due to the target materials employed.\\
Light elements are in general penalized in probing spin-in\-dependent cross sections for dark matter-nucleus elastic scattering because the expected rate scales with the square of the mass number ($\sim A^{2}$)~\cite{Kurylov03}. This disadvantage is in part mitigated by kinematics for low dark matter masses: the lighter the element, the larger the transferred momentum due to elastic scattering of dark matter particles on nuclei. \\
On the other hand, for spin-dependent interactions the expected rate is proportional to the nucleon spin coefficients ($\sim \langle S_{p/n}\rangle^2$), which differ from one isotope to the other and do not favour heavy ones~\cite{Bednyakov04}. Spin-dependent interactions can be tested only on isotopes with a nuclear ground state angular momentum \mbox{$J_N$  $\neq 0$}~\cite{Goodman1984, Ellis1991, Engel:1992bf}, therefore only a restrict\-ed number of elements fulfills this requirement.  Since the scattering kinematics remains the same as in spin-indepen\-dent interactions, it follows that certain light elements are potentially highly favoured to probe the low mass spin-\hspace{0pt plus 1pt}dependent dark matter parameter space (\unit[$\lesssim10$]{GeV/c$^2$}). Hence, the ideal target to test spin-dependent interactions should be constituted of an element with $J_N \neq 0$, $\langle S_{p/n}\rangle = 1/2$, and the lowest possible mass.\\
Currently, lithium is the lightest element contained in inorganic crystals that can be operated at cryogenic temperatures~\cite{Barinova2010, Miuchi02, Casali2013, Martinez2012, Danevich2018}. Its most abundant isotope is $^{7}$Li (92.41\% natural abundance~\cite{chimica}) with nuclear angular momentum $J_N=3/2$ and $\langle S_{p}\rangle$ close to 1/2~\cite{Bednyakov04}. For all these reasons, lithium-based crystals are very well suited to probe spin-\hspace{0pt plus 1pt}dependent interactions. Other light elements which can be competitive for spin-dependent dark matter search at low masses are hydrogen, which can be found in organic liquid scintillators~\cite{Collar2018}, and helium, which could be employed in a gaseous ionization detector.\\
We present the cryogenic operation of a prototype detector module based on a Li$_2$MoO$_4$ target crystal, which was originally developed for the CUPID-0/Mo experiment (the very same crystal is labeled \textit{LMO-3} in~\cite{LIMO1}). The results presented in this work set the most stringent limits with cryogenic detectors for spin-depen\-dent dark matter interactions with protons below \unit[1.5] {GeV/c$^2$}.

\section{Theoretical Framework} \label{sec:theo}

In the scenario typically assumed to calculate the sensitivity of a given direct detection experiment~\cite{Abdelhameed2019,Agnese17,Akerib2017,Jiang2018,Aprile2019x,Fu2016,Amole:2019,Armengaud2019}, 
a spin $1/2$ dark matter particle interacts with the nuclei of the target. At quantum level, the dark matter particle interacts with the quarks and these interactions are mediated by a heavy boson. In this framework, the differential spin-dependent elastic cross section of dark matter particles with nuclei is proportional to the non-relativistic limit of the transition amplitude between initial and final states of the axial-vector current term~\cite{Engel:1992bf, Klos:2013rwa}. 
In most dark matter scenarios the event rate is dominated by the spin-independent cross sections, which tends to explain why the majority of the existing experiments are designed to probe this type of interactions.  However, it has been shown that in some models spin-depen\-dent interactions can provide the largest contribution to the event rate~\cite{Freytsis:2010ne}. These types of scenarios strongly motivate the investigation of spin-dependent dark matter interactions. 

The differential spin-dependent cross section as function of the transferred momentum $q$ is~\cite{Jungman:1995df,Engel:1992bf, Goodman1984}:

\begin{linenomath*}
\begin{equation}
\label{diffCross}
\frac{d\sigma^{SD}}{dq^2}=\frac{8G_F^2}{(2J_N+1)v^2}S_A(q)
\end{equation}
where $G_F$ is the Fermi coupling constant, $J_N$ is the nuclear ground state angular momentum, $v$ is the dark matter particle-nucleus relative velocity, $S_A(q)$ is the axial-vector structure function. The axial-vector structure function is
\begin{equation}
S_A(q)=a^2_0\;S_{00}(q)+a_0a_1\;S_{01}(q)+a_1^2\; S_{11}(q)
\end{equation}
\end{linenomath*}
where $a_0$ and $a_1$ are the coefficients of the isoscalar-isovector parametrization of the quark axial-vector current computed among the initial and final nuclear states and $S_{ij}(q)$ are functions obtained by nuclear calculations. The value of these coefficients depends on the dark matter-quark interaction model. Even considering the maximum transferred momentum $q_{max}$, i.e. $q$ evaluated at the escape velocity $v_{esc}$ and for dark matter mass $m_\chi$ equal to the mass of the nucleus $m_N$, the axial-vector structure function for light nuclei is anyhow \mbox{$S_A(q_{max})\simeq S_A(0)$}, therefore we can safely assume the \mbox{$q^2\rightarrow0$} limit, that is equivalent to assume a form factor \mbox{$F(q)=S_A(q)/S_A(0)=1$}. In this limit,

\begin{linenomath*}
\begin{equation}
S_A(0) = \frac{(J_N+1) (2J_N+1)}{4 \pi J_N}|(a_0+a_1)\; \langle S_p\rangle + (a_0-a_1) \langle S_n \rangle |^2
\end{equation}
\end{linenomath*}
where $a_p=a_0+a_1$ and $a_n=a_0-a_1$. The fixed values $a_p=2$, $a_n=0$ and $a_p=0$, $a_n=2$ (equivalent to $a_0=a_1=1$ and $a_0=-a_1=1$) are commonly imposed for convenience and labeled as \textit{proton-only} and \textit{neutron-only} interactions, respectively. Finally, $\langle S_{p}\rangle$ and $\langle S_{n}\rangle$ are the spin matrix elements arising from the \textit{proton-only} and \textit{neutron-only} interactions. 
These spin matrix elements are a key factor to accurately calculate the cross sections for spin-dependent interactions, but, despite modern developments in the estimation of the nuclear matrix elements (e.g. including two-body currents, as in~\cite{Klos:2013rwa}), the only available literature on lithium is still the one cited in~\cite{Bednyakov04}. We will refer to the most advanced calculation ($\langle S_{p}\rangle = 0.497, $ $\langle S_{n}\rangle= 0.004$)~\cite{pacheco1989nuclear} to derive the experimental results presented in this work. The lack of updated calculations can likely be attributed to the absence of lithium-based experiments in the current panorama. In light of this work, however, we strongly encourage the computation of the nuclear matrix elements for $^{6}$Li (7.59\% natural abundance~\cite{chimica}) and $^{7}$Li (92.41\% natural abundance~\cite{chimica}) employing up-to-date techniques.\\
With this premise, we can compute the expected differential count rate for dark matter-nuclei spin-dependent interactions~\cite{Lewin:1995rx}. 
Taking into account all the numerical coefficients, the number of counts per (kg$\cdot$keV$\cdot$day) for dark matter spin\--dependent interactions is
\begin{linenomath*}
\begin{equation}
\label{rate}
\frac{dR}{dE_R} = \frac{\xi}{A}\left( \frac{\rho_0}{m_\chi}\right) 2m_T \left(\frac{J_N+1}{3 J_N}\right)\left(\frac{\langle S_{p/n}\rangle^2}{\mu^2_{p/n}}\right)\sigma^{SD}_{p/n} \eta (v_{min}) 
\end{equation}
\end{linenomath*}
where $E_R$ is the recoil energy, $A$ the target mass number, and $\xi$ a normalization factor;
$n=\rho_0/m_\chi$ is the number density of incoming particles, where $\rho_0$ is the local dark matter mass density and $m_\chi$ the dark matter mass; $m_T$ is the target mass, $\mu^2_{p/n}$ the nucleon-dark matter reduced mass, and $\sigma^{SD}_{p/n}$ the dark matter-proton/neutron cross section. Finally, $ \eta (v_{min}) $ is the mean inverse velocity in the Standard Halo Model~\cite{Freese:2012xd} where $v_{min}$ is the minimal velocity required to transfer a recoil energy \mbox{$E_R$~\cite{Lee:2013xxa}}. 

\section{Experimental setup} \label{sec:exp}
We operated a small scintillating crystal of Li$_2$MoO$_4$ with a size of (10$\times$10$\times$10) mm$^3$ and mass of \unit[2.66]{g} as cryogenic detector. The crystal constitutes the main absorber of a scintillating cryogenic calorimeter detector module~\cite{detect_concept}. This detector was operated at the Max Planck Institute (MPI) for Physics in Munich, Germany, in a dilution refrigerator Kelvinox400HA from Oxford Instruments installed in an above-ground laboratory without shielding against environmental and cosmic radiation (see~\cite{Angloher17} and references therein for details of the cryogenic infrastructure).

\begin{figure}
\centering 
\includegraphics[width=.45\textwidth]{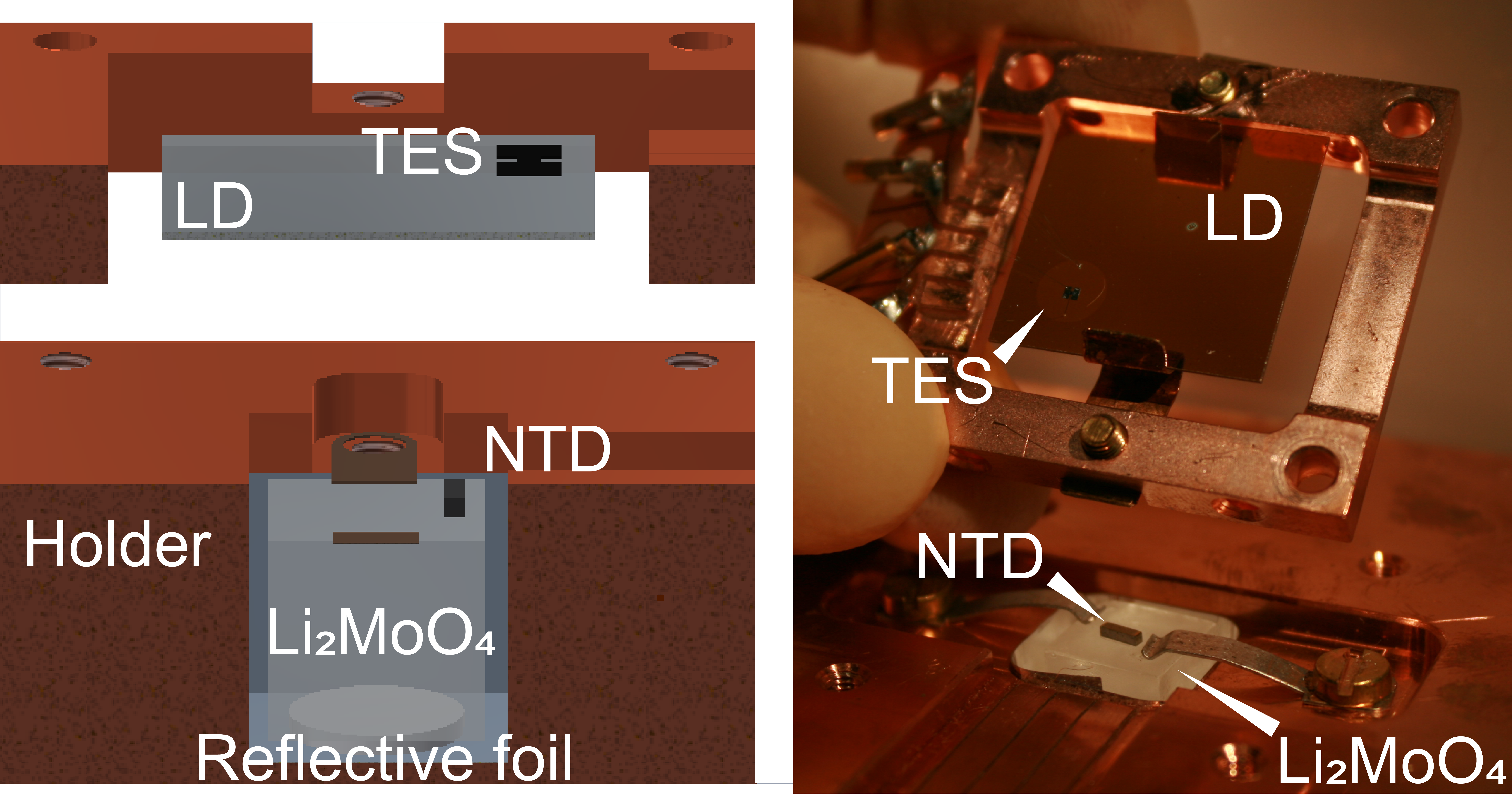}
\caption{\textbf{Left}: section view of the detector module. \textbf{Right}: picture of the detector module. The Li$_2$MoO$_4$ crystal sits on a piece of PTFE inside a reflective cavity and is held in position with two bronze clamps. One NTD of \unit[(1$\times$1$\times$3)]{mm$^3$} is glued on the top surface of the crystal and is used as thermal sensor for signal read-out. A \unit[(20$\times$20$\times$0.3)]{mm$^3$} wafer of silicon-on-sapphire is used as light absorber, its frame is fixed on top of the target crystal. The thermal sensor is a TES directly deposited on the silicon coated side of the silicon-on-sapphire plate.}
\label{fig:det_pic}
\end{figure}

\begin{figure}
\centering 
\includegraphics[width=.45\textwidth]{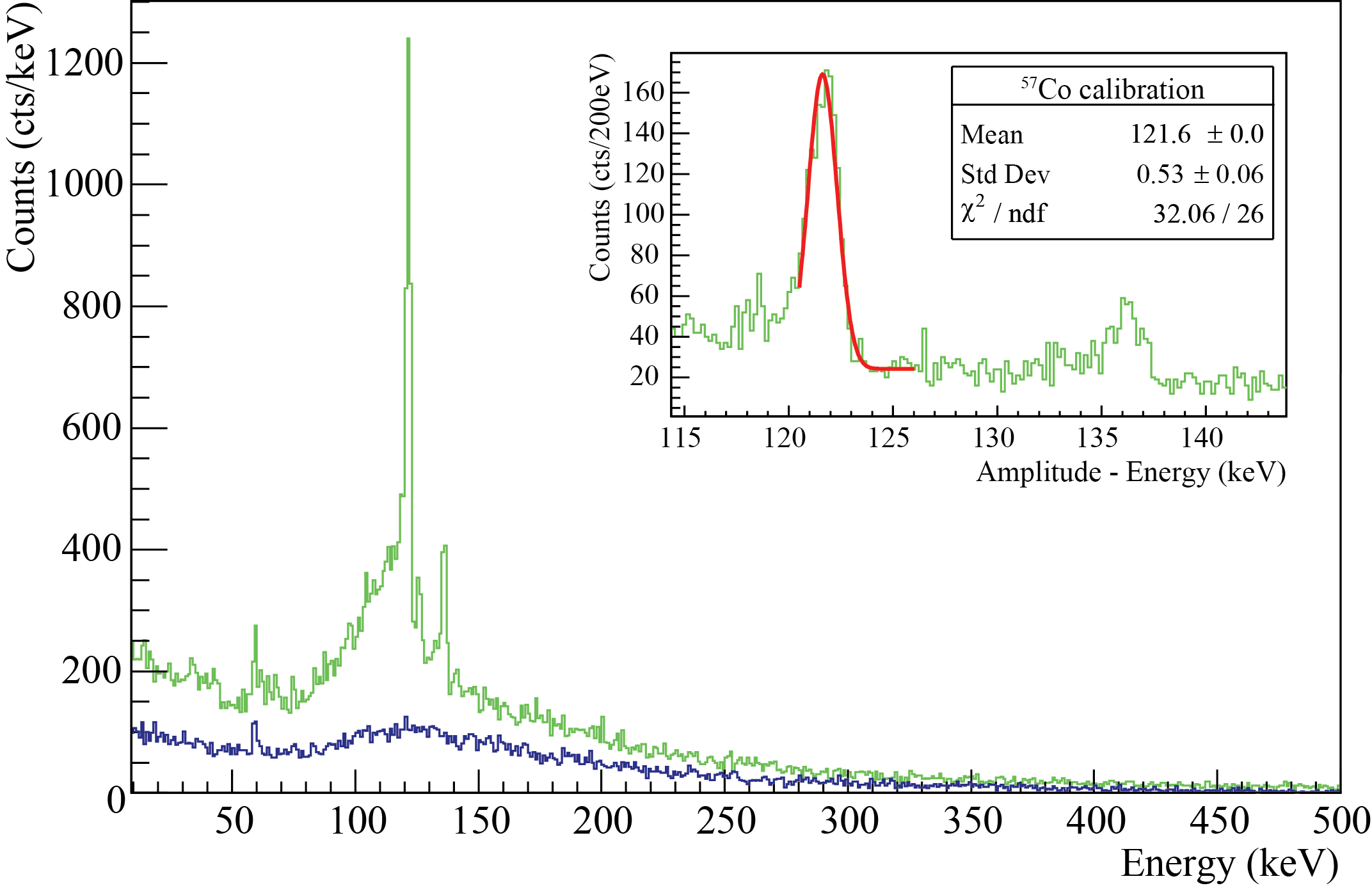}
\caption{\textbf{Green}: measured spectrum using a $^{57}$Co $\gamma$-calibration source in 3.3 hours. 
\textbf{Blue}: spectrum from 3.3 hours of background measurement. A bump peaks around 120 keV due to environmental radioactivity and a line appears at 59.5 keV due to an $^{241}$Am contamination inside the setup. 
The two prominent peaks visible only in the green plot correspond to the \unit[122]{keV} and the \unit[136]{keV} $\gamma$ rays of the $^{57}$Co source: this region of the spectrum is also visible in the inlay in the top right corner, where the fit of the \unit[122]{keV} peak is shown. 
}
\label{fig:spec}
\end{figure}

The Li$_2$MoO$_4$ crystal is held in a copper holder using bronze clamps. The internal surfaces of the holder are covered by a reflector\footnote{3M's Vikuiti\textsuperscript{TM} Enhanced Specular Reflector} to enhance the light collection efficiency. The crystal is instrumented with a \unit[(1$\times$1$\times$3)]{mm$^3$} Neutron Transmutation Doped (NTD) germanium thermistor~\cite{NTD} glu\-ed\footnote{GP 12 Allzweck-Epoxidkleber} on one surface: this sensor measures temperature variations induced by particle interactions inside the target crystal. Li$_2$MoO$_4$ is also a scintillator at cryogenic temperatures ~\cite{Barinova2010, Cardani2013}, so a fraction of the energy deposited by particle interactions is converted into scintillation light. The light is detected using a CRESST-III light detector (LD)~\cite{Rothe2018}, made of a \unit[(20$\times$20$\times$0.3)]{mm$^3$} sapphire wafer coated on one face with a 1 $\mu$m thick silicon layer (Silicon-on-Sapphire, SOS) where a Transition Edge Sensor (TES), used as thermal sensor, is deposited. The sapphire side of the LD is facing the upper side of the Li$_2$MoO$_4$ crystal. Electrical and thermal connections are provided to the LD and the NTD via \unit[25]{$\mu$m} diameter aluminum and gold bond wires, respectively. The temperature of the NTD is read out by measuring the voltage drop of the sensor with a commercial differential voltage amplifier~\footnote{Stanford Research System https://www.thinksrs.com/products/sr560.html} while applying a constant bias current through the NTD. The readout of the LD, instead, is obtained with a commercial SQUID~\footnote{Applied Physics System model 581 DC SQUID} system, combined with a CRESST-like detector control system~\cite{ANGLOHER2009}. An $^{55}$Fe X-ray source with an activity of \unit[0.055]{Bq} was placed about \unit[0.5]{cm} away from the light detector to calibrate its energy response. \\ The two detectors were combined to constitute a detector module (Figure \ref{fig:det_pic}): this module was then mechanically and thermally connected to the coldest point of the  dilution refrigerator, which retained a temperature of $\sim$\unit[10]{mK} during the whole data collection. This temperature is optimal for the NTD operation, but not for the LD. This particular TES, in fact, showed a critical temperature of \mbox{$T_{c}=\unit[22]{mK}$}. Hence, the operating point of the LD had to be stabilized around $T_{c}$ using a heater made of a thin gold film directly deposited in proximity to the TES. \\
Three measurement campaigns were performed: a gamma calibration, a neutron calibration, and a background measurement. First, a $^{57}$Co $\gamma$-source was placed outside the cryogenic system for gamma calibration, which resulted in two visible lines in the spectrum at \unit[122]{keV} and at \unit[136]{keV}. Then, an AmBe source was placed in a similar position for neutron calibration. Finally, we removed the source to collect \unit[14.77]{hours} of gross background data before the end of the measurement. \\ 
The two spectra computed in the \unit[1--500]{keV} energy range for 3.3 hours each of stable phonon detector operations are shown in Figure \ref{fig:spec}. The phonon detector shows a consistent pulse shape up to MeV energy scale. The detector response is calibrated on the \mbox{\unit[122]{keV}} and \unit[136]{keV} peaks using a linear regression with the y-intercept constrained to 0 and the first order coefficient as a free parameter. We also observe a third peak due to an $^{241}$Am contamination inside the set-up in all three measurement campaigns. Using the calibration factor obtained with the fit, the $^{241}$Am $\gamma$-line appears at \unit[(59.5$\pm$0.2)]{keV}, which matches the expected value of \unit[59.54]{keV}~\cite{americium}. For this reason and given the response function of an NTD~\cite{NTD}, we can safely assume  that the energy response is linear in the \unit[0]-\unit[136]{keV} range. After calibration, we quote the response of the NTD as \unit[(848$\pm$11)]{nV/keV}. The energy resolution at zero energy, also denoted as baseline resolution, is $\sigma_{baseline}=$(0.174$\pm$0.006) keV and the energy resolution at \unit[122]{keV} is $\sigma_{\gamma}=$(0.53$\pm$0.06) keV. We also observe the \unit[4.78]{MeV} thermal neutron capture peak of $^6$Li which has a resolution of $\sigma_{n_{cap}}=\hspace{0pt plus 1pt}$(2.36$\pm$0.14) keV. The aforementioned energy resolutions are obtained via a Gaussian fit where standard deviation, center position, and amplitude are free parameters. 
The measured background rate is 2.37$\times 10^4$ counts/(keV$\cdot$kg$\cdot$day) in the \unit[1--200]{keV} range.
The LD is calibrated on the \unit[5.89]{keV} peak of $^{55}$Fe and has a baseline resolution $\sigma^{LD}_{baseline}$=\unit[(5.90$\pm$0.13)]{eV}. The detector module shows a light yield (LY) for $\beta$ and $\gamma$ particles of (0.32$\pm$\hspace{0pt plus 1pt}0.01) keV/\hspace{0pt plus 1pt}MeV. The LY was computed as the ratio of the scintillation light detected in the LD converted in energy over the total energy deposited in the main absorber. The value we obtained is lower than previous cryogenic measurements with a similar crystal~\cite{LIMO1, Mancuso2016}: we attribute this discrepancy to the different experimental setup (i.e. different LD and different geometry).
The resulting quenching factors~\cite{Tretyak2009} are 0.205$\pm$\hspace{0pt plus 1pt}0.007 for $\alpha$ particles and 0.124$\pm$0.012 for nuclear recoils induced by neutrons as seen during the neutron calibration, in agreement with the literature~\cite{LIMO1}. 

\begin{figure}
\centering 
\includegraphics[width=.45\textwidth]{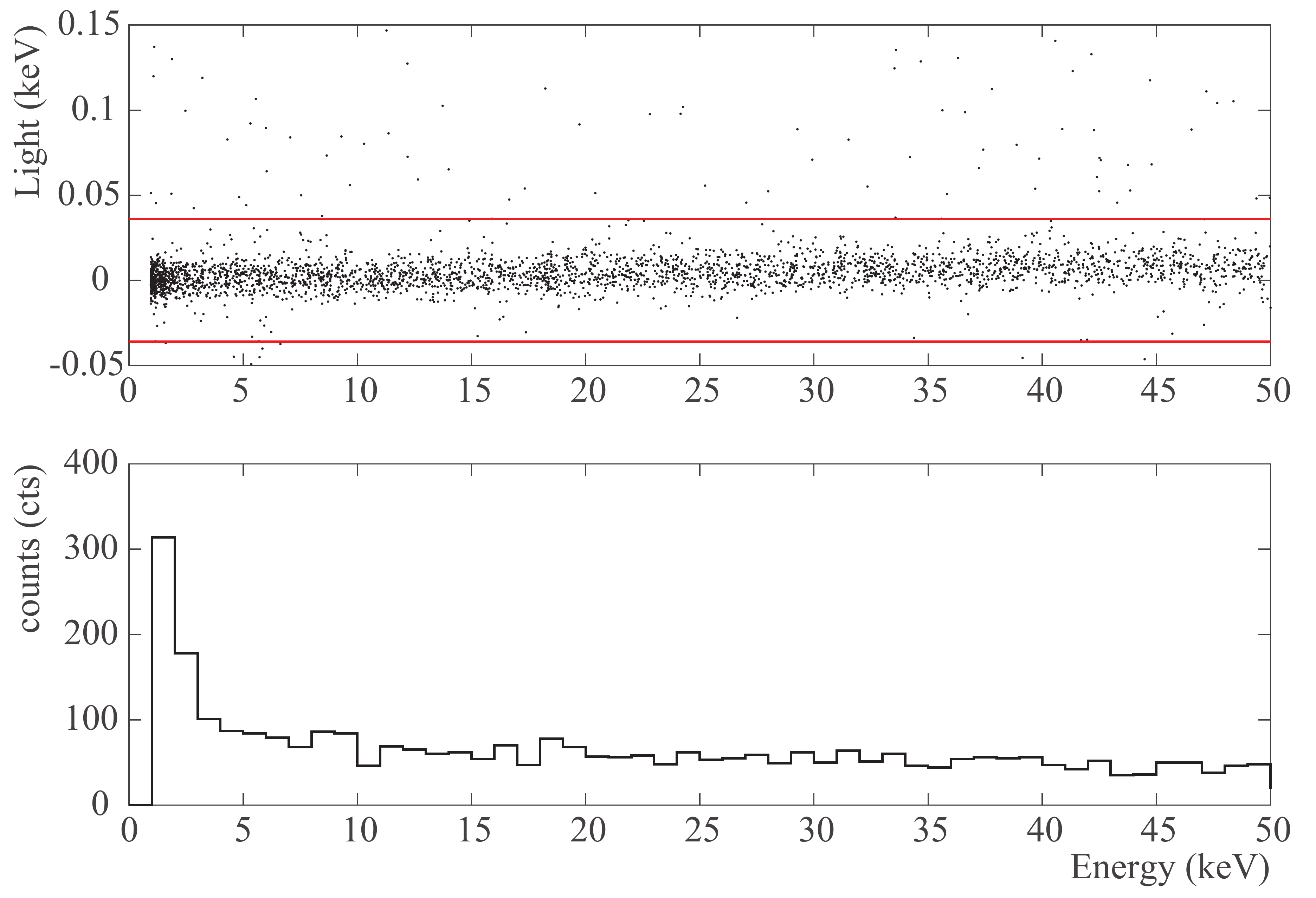}
\caption{\textbf{Top}: the light measured in coincidence by the LD (y-axis) is displayed against the energy deposited in the Li$_2$MoO$_4$ crystal (x-axis) in the ROI (\unit[1-50]{keV}). The two lines in solid red correspond to the values chosen for the anti-coincidence cut: the events which fall inside the two lines are accepted for the dark matter analysis. The rejected events show an excessive light signal, which cannot be attributed to single particle hits in the main absorber. 
\textbf{Bottom}: measured energy spectrum of the selected events. Those events can mainly be attributed to low energy $\gamma$ rays.}
\label{fig:LY}
\end{figure}

\section{Dark Matter Results} 
\label{sec:DMResults}

The spin-dependent dark matter limits we present were calculated using the background measurement dataset. The results obtained
should be seen as an evidence of the high potential of lithium-based crystals, rather than a conclusive outcome. For this very reason we decided to adopt a conservative approach for the data analysis and to collect only a few hours of background data. There would be no major benefit to aim for a longer data taking and for a more stringent data selection, since we are intrinsically limited in a non-shielded above-ground laboratory.

The energy region of interest we chose to compute our dark matter results is ranging from threshold to \unit[50]{keV}.
Due to the poor LY, in this energy range we cannot perform a particle identification analysis. Thus, the light signal is used only as a veto for muons and events originating from the materials surrounding the crystal. 
We expect dark matter particles to directly interact only with the Li$_2$MoO$_4$ crystal, but never with both the crystal and the light detector simultaneously. For this reason we define a region of interest (ROI) in the two dimensional space described by the energy deposited in the crystal on the x-axis and the energy deposited in the LD on the y-axis (see Figure~\ref{fig:LY}, top). The ROI is defined on the x-axis by the energy region of interest. On the y-axis, instead, we set the maximum and the minimum values as \textit{C} and \textit{-C} respectively, where \textit{C} is defined as
\begin{linenomath*}
\begin{equation}
\label{cut} C = L_{max} + 2 \cdot \sigma^{LD}_{\gamma}= 39.2~\textrm{eV} \end{equation}
\end{linenomath*}
This definition takes into account the maximum scintillation expected in the energy region of interest $L_{max}$ and the energy resolution of the light detector $\sigma^{LD}_{\gamma}$. \\
$L_{max}$ is simply obtained by the multiplication of the LY with the maximum value in the energy region of interest: 
\begin{linenomath*}
\begin{equation}
\label{ly} L_{max}= LY \cdot 50\textrm{keV} = 12.8\textrm{eV}\end{equation}
\end{linenomath*}
Finally, the energy resolution \mbox{$\sigma^{LD}_{\gamma}=$\unit[(10.0$\pm$1.6)]{eV}} is computed using the peak resulting from the scintillation generated by the absorption of \unit[122]{keV} $\gamma$ rays in the crystal during the $^{57}$Co calibration.
All the events falling inside the ROI are accepted for the dark matter analysis (see Figure~\ref{fig:LY}, bottom) without further data selection. \\The events falling outside the ROI are contributing to the dead time, hence the effective measurement time is reduced to 9.68 hours, which corresponds to a $^7$Li exposure of \unit[7.91\hspace{0pt plus 1pt}$\times$10$^{-5}$]{kg$\cdot$day}.
The energy threshold of 0.932$\pm$0.012 keV has been determined according to the procedure described in~\cite{trigger}. Given the background induced by our setup, we set the threshold allowing a noise trigger rate (the rate of events caused by noise oscillation) of \unit[$1\times10^4$]\hspace{0pt plus 1pt}{counts/(keV$\cdot$kg\hspace{0pt plus 1pt}$\cdot$day)}, which leads to a contribution in the first bin spectrum of approximately 10$\%$ of the total triggered events. The method we applied is valid in low-rate measurement conditions, a requirement we do not satisfy, therefore a higher trigger pede\-stal is expected manly due to pileup.

\begin{figure}[t]
  \includegraphics[width=.5\textwidth]{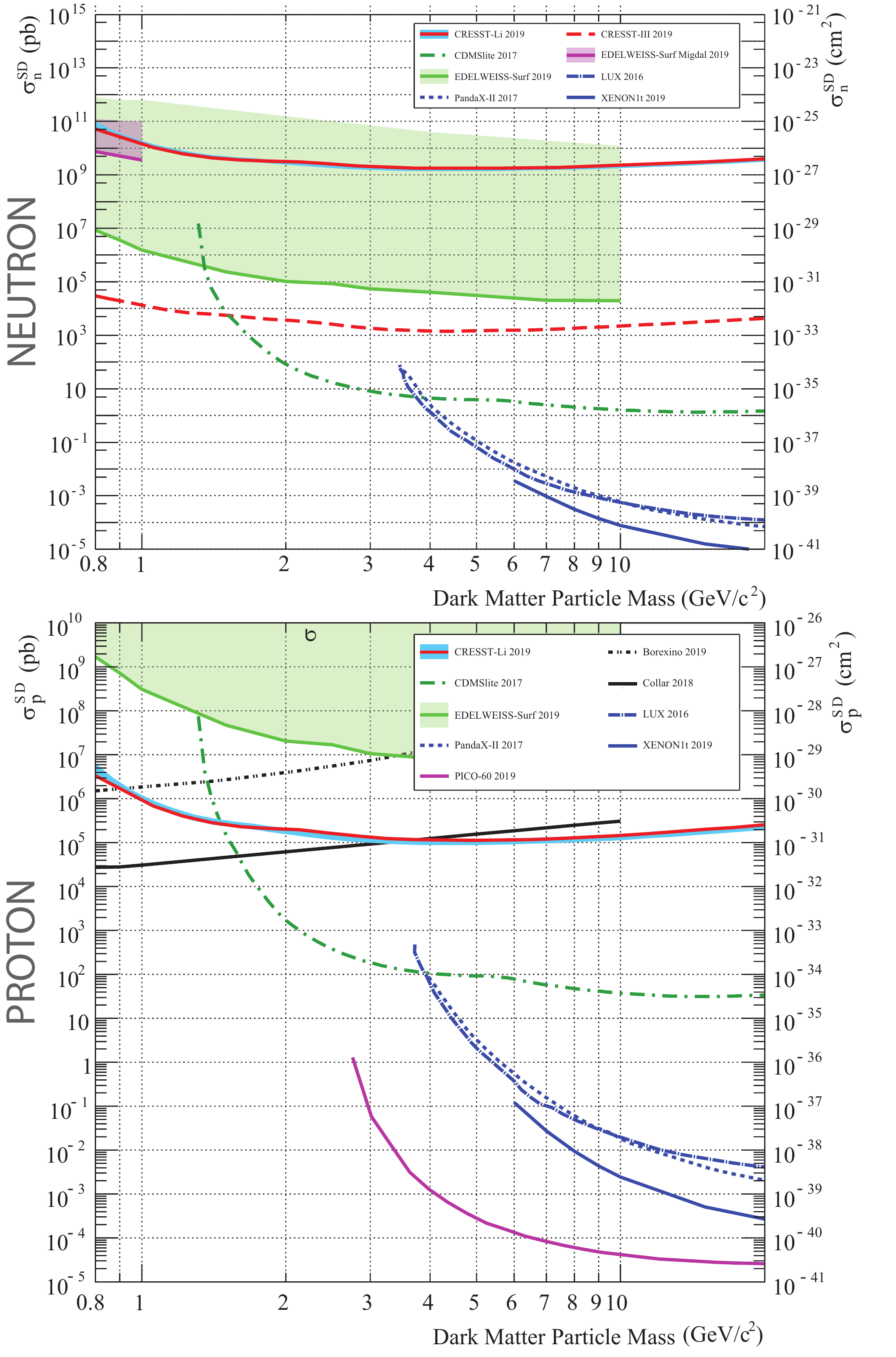}
\caption{%
\textbf{Top:} Exclusion limit obtained for \textit{neutron-only} spin-dependent interactions of dark matter particles with Standard Model particles. The cross section for this kind of interactions is shown on the y-axis (pb on the left, cm$^2$ on the right), while the dark matter particle mass is on the x-axis. The result of this work with $^7$Li is drawn in solid red with the two-sigma band resulting from statistical uncertainty in solid blue: we reach \unit[1.06$\cdot$10$^{-26}$]{cm$^2$} at \unit[1]{GeV/c$^2$}. In dashed red we show the CRESST-III~\cite{Abdelhameed2019} limit using $^{17}$O. For comparison, we show limits derived by other direct detection experiments: EDELWEISS~\cite{Armengaud2019} and CDMSlite~\cite{Agnese17} using $^{73}$Ge; LUX~\cite{Akerib2017}, PandaX-II~\cite{Fu2016}, and XENON1T~\cite{Aprile2019x} using $^{129}$Xe+$^{131}$Xe (see legend).
\textbf{Bottom:} Same, but for \textit{proton-only} spin-dependent interactions. Our result with $^7$Li is depicted in solid red with with the two-sigma band in solid blue, reaching \unit[6.88$\cdot$10$^{-31}$]{cm$^2$} at \unit[1]{GeV/c$^2$}. Additionally, we plot limits from other experiments: CDMSlite~\cite{Agnese17} and EDELWEISS~\cite{Armengaud2019} with $^{73}$Ge; LUX~\cite{Akerib2017}, XENON1T~\cite{Aprile2019x}, and PandaX-II~\cite{Fu2016} with $^{129}$Xe+$^{131}$Xe; PICO-60 with $^{19}$F~\cite{Amole:2019}; Collar~\cite{Collar2018} with $^{1}$H. Finally, we plot in dotted black a constraint from Borexino data derived in~\cite{Bringmann2018}.
}
  \label{fig:limitProton}
\end{figure}

We treat all events in the energy range between 0.932 keV and \unit[50]{keV} as potential signal events, not performing any background subtraction and we conservatively calculate exclusion limits on spin-dependent interactions of dark matter particles with nuclei using Yellin's optimal interval method~\cite{Yellin02, Yellin08} valid for \textit{proton-only} interactions and for \textit{neutr\-on-only} interactions, as discussed in the theoretical fra\-mework presented before. 
For the calculation of the exclusion limits we adopt the standard dark matter halo model, which assumes a dark matter halo with a Maxwellian velocity distribution and a local dark matter density of $\rho_\text{DM} = \unit[0.3]{GeV/ \- ( c^{2}\cdot cm^{3})}$~\cite{Salucci2010}. We also assume $v_\text{esc} = \unit[544]{km/s}$ for the galactic escape velocity~\cite{Smith2006} and $v_\odot = \unit[220]{km/s}$ for the solar orbit velocity~\cite{Kerr1986}. 
We tested the trigger efficiency generating a known flat energy spectrum of events. Each event is generated superimposing the ideal detector response, scaled to match the amplitude of simulated energy, on the recorded data. The simulated data is then processed with the same algorithm used for the real data. The fraction of survived events over the total simulated events at each energy represents the trigger efficiency, which was included in the calculation of the exclusion limits. 
Figure \ref{fig:limitProton} shows the results obtained for \textit{proton-only} and \textit{neutron-only} interactions and the associated two-sigma statistical uncertainty. These results are extremely competitive with other spin-dependent direct dark matter searches for very light dark matter particles masses, especially in the sub-GeV/c$^2$ regime.
For dark matter masses  \unit[$\gtrsim1.5$]{GeV/c$^2$} our results are not competitive with other direct search experiments, such as PICO-60~\cite{Amole:2019}, CDMSlite~\cite{Agnese17},  LUX~\cite{Akerib2017}, CDEX-10~\cite{Jiang2018}, XENON1T~\cite{Aprile2019x}, PandaX-II \cite{Fu2016}, due to the small exposure and the substantially higher background level, mainly caused by the ab\-ove-ground operation in a non-shielded environment. Considering these sub-optimal conditions and reversing the argument, these results convincingly show the benefit of a comparable low threshold combined with a light target nucleus. The versatility to change the target material is a key feature of cryogenic detectors in general and CRESST-like readout in particular. This clearly yields the prospect of a quick advancement of sensitivities in the low-mass dark matter sector for spin-dependent interactions in the near future. 

\section{Conclusions}
We have successfully operated a scintillating cryogenic detector based on 2.66 g of Li$_2$MoO$_4$ target crystal at the Max Planck Institute (MPI) for Physics in Munich, Germany.  After testing the detector response in presence of a neutron source and a $^{57}$Co $\gamma$ rays source, we performed a background measurement lasting 9.68 hours of effective time, achieving an energy threshold of \unit[(0.932$\pm$0.012)]{keV}. This measurement sets the cornerstone for the use of lithium-based crystals in the low-mass spin-dependent dark matter sector and shows that it is possible to obtain extremely competitive results for masses below \unit[1.5]{GeV/c$^2$} even using a non-optimal phonon detector in a high background experimental setup.\\
We plan future measurements with lithium-based crystals, a CRESST-like phonon detector, and an underground experimental setup which could drastically boost the sensitivity with respect to this work.

\begin{acknowledgements}
This work has been supported through the DFG by the SFB1258 and the Excellence Cluster Universe, and by the BMBF 05A17WO4. 
\end{acknowledgements}

\bibliographystyle{h-physrev}      

\end{document}